\documentclass[]{aa}  

\usepackage{graphicx}
\usepackage{txfonts}
\usepackage{lscape}
\usepackage{listings}
\usepackage{color}
\usepackage{graphicx}
\usepackage{caption}
\usepackage{subcaption}
\usepackage{grffile}
\usepackage{hyperref}
\definecolor{mygreen}{rgb}{0,0.6,0}
\definecolor{mygray}{rgb}{0.5,0.5,0.5}
\definecolor{mymauve}{rgb}{0.58,0,0.82}
\definecolor{CJosefa}{rgb}{1.0, 0.65, 0.0} 
\definecolor{RKblue}{rgb}{0.6, 0.0, 1.0} 

\usepackage{soul,xcolor,color}
\usepackage{rotating}

\begin{document} 

\setstcolor{magenta}

 \title{Orion revisited}

   \subtitle{III. The Orion Belt population}

   \author{K. Kubiak
          \inst{\ref{vienna}}
          \and
          J. Alves \inst{\ref{vienna}}
          \and
          H. Bouy\inst{\ref{herve}}
          \and 
          L. M. Sarro\inst{\ref{luis}} 
           \and 
          J. Ascenso\inst{\ref{joana1}, \ref{porto}}
          \and
          A. Burkert\inst{\ref{USM}, \ref{MPE}}
          \and 
          J. Forbrich\inst{\ref{vienna}} 
          \and
 	J. Gro{\ss}schedl\inst{\ref{vienna}}
          \and
          A. Hacar\inst{\ref{vienna}}  
          \and 
          B.~Hasenberger\inst{\ref{vienna}}
          \and 
          M. Lombardi\inst{\ref{italy}}
          \and 
          S. Meingast\inst{\ref{vienna}}
	 \and
	R. K\"ohler\inst{\ref{innsbruck},\ref{vienna}}
          \and 
          P. S. Teixeira\inst{\ref{vienna}}
          }

   \institute{ Department of Astrophysics, University of Vienna,
              T\"urkenschanzstrasse 17, A-1180 Vienna \label{vienna}\\
              \email{karolina.kubiak@gmail.com }
        	     \and
             Center for Astrobiology (INTA-CSIC), Camino Bajo del Castillo S/N,  E-28692 Villanueva de la Ca\~nada, Madrid \label{herve}
             \and 
             University of Milan, Department of Physics, via Celoria 16, 20133, Milan, Italy\label{italy}
             \and 
             CENTRA, Instituto Superior Tecnico, Universidade de Lisboa, Av. Rovisco Pais 1, 1049-001, Lisbon, Portugal\label{joana1}
             \and 
             Universidade do Porto, Departamento de Engenharia Fisica da Faculdade de Engenharia, Rua Dr. Roberto Frias, s/n, 4200-465, Porto, Portugal\label{porto}
             \and 
             Dpto. de Inteligencia Artificial, ETSI Informatica, UNED, Juan del Rosal, 16, 28040, Madrid, Spain\label{luis}
             \and
             Institut f\"ur Astro- und Teilchenphysik, Universit\"at Innsbruck, Technikerstr. 25/8, 6020 Innsbruck, Austria\label{innsbruck}
             \and
             Universit\"ats-Sternwarte Ludwig-Maximilians-Universit\"at (USM),     Scheinerstr. 1, M\"unchen, D-81679, Germany \label{USM}
             \and
            Max-Planck-Institut f\"ur extraterrestrische Physik (MPE), Giessenbachstr.1, D-85748 Garching, Germany\label{MPE}    
             }

   \date{\today}

 
  \abstract
   {}
 {This paper continues our study of the foreground population to the Orion molecular clouds. The goal is to characterize the foreground population north of NGC 1981 and to investigate the star formation history in the large Orion star-forming region. We focus on a region covering about 25 square degrees, centered on the $\epsilon$ Orionis supergiant (HD 37128, B0\,Ia) and covering the Orion Belt asterism.}
   {We used a combination of optical (SDSS) and near-infrared (2MASS) data, informed by X-ray (\textit{XMM-Newton}) and mid-infrared (WISE) data, to construct a suite of color-color and color-magnitude diagrams for all available sources. We then applied a new statistical multiband technique to isolate a previously unknown stellar population in this region.}
      {We identify a rich and well-defined stellar population in the surveyed region that has about 2\,000 objects that are mostly M stars.  We infer the age for this new population to be at least 5\, Myr and likely $\sim10$\,Myr and estimate a total of about 2\,500 members, assuming a normal IMF. This new population, which we call the Orion Belt population, is essentially extinction-free, disk-free, and its spatial distribution is roughly centered near $\epsilon$ Ori, although substructure is clearly present.}
        {The Orion Belt population is likely the low-mass counterpart to the Ori OB Ib subgroup. Although our results do not rule out Blaauw's sequential star formation scenario for Orion, we argue that the recently proposed blue streams scenario  provides a better framework on which one can explain the Orion star formation region as a whole.  We speculate that the Orion Belt population could represent the evolved counterpart of an Orion nebula-like cluster.}

\keywords{Stars: formation - Stars: late-type - Stars: pre-main sequence - ISM: clouds }

\maketitle

\section{Introduction}

\begin{figure}[h!]
        \centering
      \includegraphics[width=\hsize]{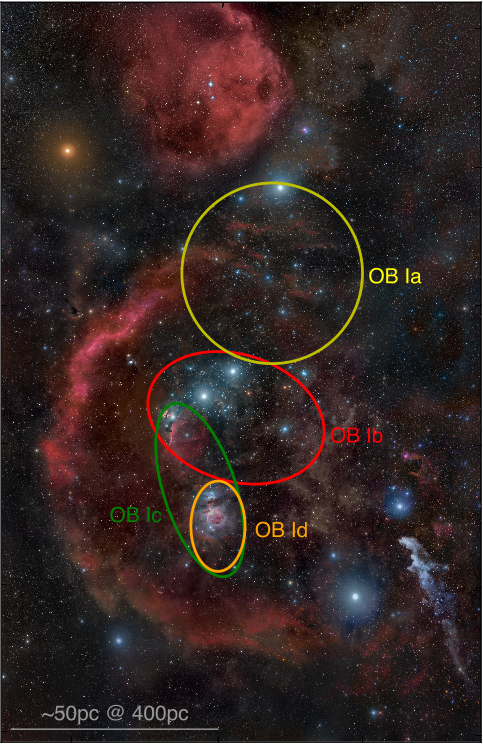}
                \caption{Widefield image of the Orion OB\,I stellar associations as described by A. \cite{Blaauw} and revised in \cite{Bally2008}. North is up, and east is left. Background image: R. Bernal Andreo, \url{www.deepskycolors.com}.}   \label{fig:Orion}
\end{figure}

The Orion star formation complex is the closest massive star-forming region to the Sun and has generated about $10^4$ low- and high-mass stars for at least the last $\sim\,12$ Myr (e.g., \cite{Blaauw}; \cite{Brown1994}; \cite{Bally2008}; \cite{Muench2008}; \cite{Briceno2008}). The entire region, also known as the Orion OB\,I association, covers an area of approximately 10\degr\,$\times$\,20\degr\ on the sky and harbors a half dozen subgroups containing well-known OB stars and giant molecular clouds (see Fig.~\ref{fig:Orion}). The proximity of the region ($\sim$\,400\;pc; \cite{Hirota2007}; \cite{Menten2007}; \cite{Sandstrom2007}; \cite{Bally2008}) makes it one of the most significant star formation laboratories in astronomy. Indeed, much has been learned in Orion about star formation, for example, clues to the evolution and destruction of clouds, the physics and dynamics of the interstellar medium (ISM), and the role that OB associations and high-mass stars play in the cycling of gas between various phases of the ISM. It is very remarkable, although understandable given its size, that for a  region of such fundamental importance most attention has been devoted to the embedded and dusty stellar populations (age $\leq$3 Myr) emerging from the molecular clouds complexes Orion A and Orion B \cite[e.g.,][]{ELada_1991, Allen, Megeath2012, Gutermuth09, daRio10, Spezzi15}, whilst only  a few studies have tackled the Orion star-forming region as a whole. 

The Orion OB\,I association \citep{Blaauw} is composed of several stellar subgroups of different ages, gas, and dust amount. Blaauw divided Orion's association into four groups. Figure \ref{fig:Orion} presents a widefield image of the Orion Constellation superimposed with ellipses denoting the approximate boundaries of these four OB\,I subgroups. These groups appear to show a spatial-temporal relation that is suggestive of a sequence of star formation events, from dust free Ia subgroup to still dust embedded Id. This led  \cite{Blaauw} to propose a sequential star formation scenario, where a previous generation of stars is responsible for the formation of a new one via positive feedback; this idea was later quantified by \cite{Elmegreen_Lada} and has remained very popular in the literature. 

Although there are differences in the estimated ages or exact sizes of the various groups, most of the published  works in the region agree that the Orion OB\,Ia group toward the north is the oldest with an age of $\sim$\,8-10\,Myr (\citealt{Bally2008}, with a distance $\sim$\,350\,pc) or even 12\,Myr as originally proposed by \cite{Blaauw}. This group is also dust free. The OB\,Ib subgroup, containing the stars around the Orion Belt asterism, is located at a distance of $\sim$\,400\,pc \citep{Bally2008}. This subgroup has an estimated age of $\sim$\,3-6\,Myr \citep{Bally2008} or $\sim$\,1.7\,Myr \citep{Brown1994}, although the lower estimate is inconsistent with the age of the three supergiants ($\zeta$ Ori, $\epsilon$ Ori, and $\delta$ Ori) that form the naked-eye Belt.  According to their spectral types, these three stars must be at least 5 Myrs old. The 3-6\,Myr old OBI\,c subgroup consists of stars around the Sword (about 4$^{\circ}$ below the Belt asterism). The older stars in the OB\,Ic are superimposed on the much younger and still embedded subgroup OB\,Id, which is associated with the Orion nebula and including the Trapezium stars, M43, NGC 1977, and the OMC1, 2, and 3 regions in the integral shaped filament along with the northern part of the Orion A molecular cloud (age $<$ 2\,Myr, d$\sim$\,420pc, \cite{Bally2008}). Although initially each subpopulation was assumed to be a distinct episode in a large star formation event, it was early realized that the subgroups are partially superimposed along our line of sight and several authors have described the boundaries between subgroups, their characteristics, and some discrepancies with the sequential star formation scenario \cite[e.g.,][]{Warren1978,deGeus1990,Brown1994,Gomez1998}. Unfortunately, the three-dimensional arrangement of star-forming regions, in particular massive ones, is far from simple and is essentially unknown for any massive star-forming region given the current distance accuracies.

Recently, \cite{revisited1} and \cite{revisited2} presented evidence for a young and massive foreground population ($\sim5$ Myr) that is detached from the Orion A cloud but seen in projection toward it. They argue that this foreground population was  formed about $4-5$ Myr ago in a different, but perhaps related, event in the larger Orion star formation complex and not in the existing Orion A molecular cloud like, for example, the Orion nebula cluster. This foreground population includes in part Blaauw's OB Ic population, but does not include the younger $\sigma$-Ori cluster, as suggested in \cite{Bally2008}. An intriguing result of their study was that the Orion A foreground population seemed to extend to the north, toward OB Ib, beyond the limits of their survey.  This was later confirmed in \cite{Meingast2016} in their ESO-VISTA near-infrared (NIR) imaging of the entire Orion A cloud, who found a southern boundary to the foreground population, but not an obvious boundary toward the north. This raises the question of how well Ori OB\,Ic and Ib are separated spatially, if at all, and how they fit in a sequential star formation scenario. 

In a related study, \cite{streams} revisited the Hipparcos catalog and studied the spatial distribution in 3D of OB stars that are closer than 500\,pc from the Sun. Their analysis reveals that massive OB stars form large-scale structures that are well defined and elongated, which they refer to as ``blue streams''. The spatial coherence of these blue streams, and the monotonic age sequence over hundreds of parsecs, suggest that they are made of young stars. The two main blue streams are the Sco-CMa stream, including the Sco-Cen association, and the surprising Orion stream, originating in the Orion clouds and extending to regions as close to Earth as $\sim200$ pc, but likely even closer. In this scenario, the foreground population presented in \cite{revisited1} and \cite{revisited2} could be part of the Orion blue stream. Given this new scenario for the distribution of young stars in the local neighborhood, and in particular the realization that the Orion OB\,I association may be part of the Orion stream, there is a clear need to gather more information about the region on larger scales than collected previously and for regions further away from the molecular clouds. 

The main goal of this paper is to extend the work of \cite{revisited1} and \cite{revisited2} to the north and,  (i) further investigate the extent of the young foreground population presented, (ii) investigate the relation between OB Ic and Ib populations, and (iii) contribute to the reconstruction of the star formation history of the Orion complex. We focus our study on Blaauw's subgroup OB\,Ib by studying almost 30 square degrees of sky centered on Orion's Belt. Within the limitations of our data, we compare the recently proposed Orion blue stream scenario with that of Blaauw's classical sequential star formation. 

The overdensity of blue massive stars in the Orion Belt region was first pointed out in Galileo's Sidereus Nuncius in 1609 as an example of how the telescope could resolve stars that are not visible by the human eye. The stellar overdensity was also recognized in 1931 by Swedish astronomer Per Collinder in his catalog of open clusters \citep{Collinder}. He distinguished the Orion's Belt asterism, comprised of the three famously aligned bright stars: Alnitak ($\zeta$ Ori, HD\,37742J, O9.7\,Ib+B0\,III), Alnilam ($\epsilon$ Ori, HD\,37128, B0\,Ia), and Mintaka ($\delta$ Ori, HD\,36486, B0\,III +O9\,V) as Collinder 70 (Col 70). Still, and even though it is immediately recognizable to the naked eye, the Orion Belt stellar population is paradoxically poorly known. \cite{Caballero_Solano2008} observed two circular areas of 45 arcmin radius each, centered on  Alnilam and Mintaka and found  136 low-mass stars  displaying features of extreme youth, and a total of 289 young stars in the surveyed area. They concluded that the two regions  could be analogs to the $\sigma$ Ori cluster, but more massive,  extended, and slightly older.

Since the seminal work of Blaauw, it has been suggested that the age, distance, and radial velocity of the stellar components of subgroup OB\,Ib may not be consistent with a simple sequential star formation scenario (e.g., \cite{Hardie1964}; \cite{Warren1978}; \cite{Guetter1981}; \citealt{Gieseking1983}). In particular, the eastern part of the subgroup, which includes Alnitak, the Horsehead Nebula, the Flame Nebula (associated with NGC\,2024 in the Orion B cloud), and the H\,II region IC\,434, would be the farthest and youngest subgroup. The fourth brightest star in Orion's Belt is $\sigma$ Ori (48\,Ori, HD\,37468, O\,9.5\,V),  the brightest source in the well-studied $\sigma$ Orionis Cluster \citep{Walter1997}, which has been assigned to OB\,Ib based on its spatial proximity. Still, two solid cases can be made against $\sigma$ Ori belonging to OB\,Ib. These two cases, which are discussed later in this paper, are, first, the age of the $\sigma$-Ori cluster (3 Myr; \cite{Caballero2008}) is younger than most of the stars in the Belt region  and, second, the radial velocity of stars toward the cluster shows that the young $\sigma$-Ori cluster consists of two spatially superimposed components that are kinematically separated by 7 km/s in radial velocity \citep{Jeffries2006}. In the review of \cite{Bally2008}  the $\sigma$-Ori cluster appears as member of OB\,Ic, as in Figure~\ref{fig:Orion}.

This paper is organized as follows. The next section briefly describes the data used in this study. Section \ref{sec:results} and \ref{sec:properties} present our results, centering on the discovery of a large population of young stars around $\epsilon$ Ori. In Section \ref{sec:discussion} we discuss our results and we summarize them in Section \ref{sec:summary}. 

\section{Data}\label{sec:data}
\subsection{Survey field}
\begin{figure}[h!]
\begin{center}
\includegraphics[width=\hsize]{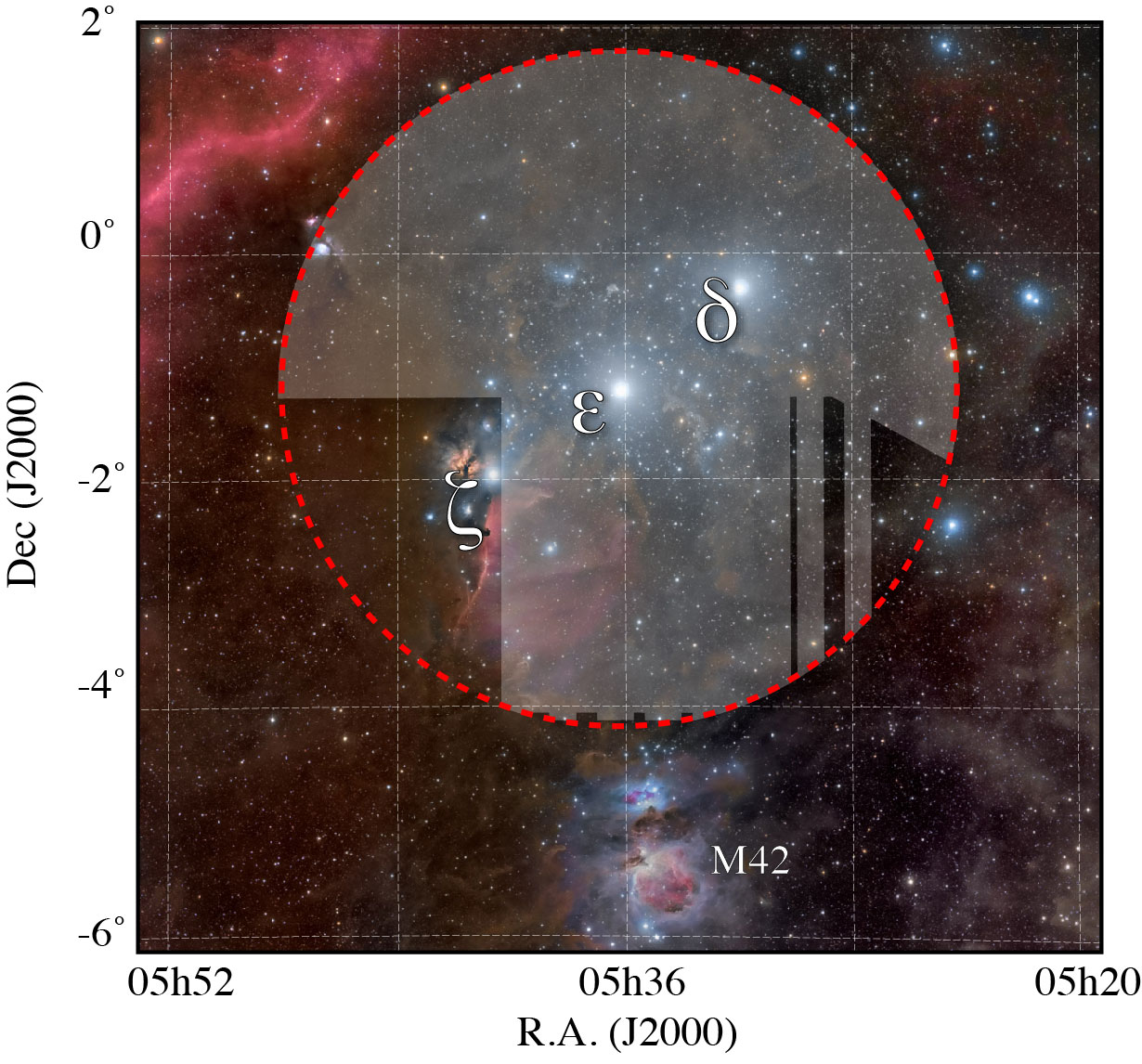}
\caption{Coverage of the study. The selected SDSS survey coverage is represented by the gray area. The 2MASS survey, an all-sky survey, is available for the entire region of study (red dotted circle). Background image: R. Bernal Andreo, \url{www.deepskycolors.com}. \\}
\label{fig:surveys}
\end{center}
\end{figure}

We first retrieved and cross-matched all the sources from the Two Micron All-Sky Survey \citep{2MASS} and Sloan Digital Sky Survey DR12 (\citealt{SDSS}) catalogs located within a radius of 3\,degrees centered around $\epsilon$ Ori at $(l,b)$=$(205.21, -17.24)\degr$. The size and location of our field were chosen to achieve the best coverage of Ori\,Ib in both catalogs. The 2MASS catalog homogeneously covers the entire area of interest, while the SDSS DR12 catalog is missing a significant fraction in the southern half. Figure \ref{fig:surveys} shows the  coverage of the different surveys used in this study over a photograph of the region. It includes, by design, most of the $\sigma$-Ori, and a significant fraction of the NGC~1980/NGC~1981 area surveyed in \cite{revisited2}.  All in all, about 21\% of the $\sim 28\,\rm deg^2$  surveyed area is incomplete. A total of 200\,497 2MASS and 909\,619 SDSS sources were found in the corresponding area. Of these,  189\,620 sources appear in both surveys.

In an effort to compile the most complete data set in terms of spatial and wavelength coverage, we then collected complementary photometry from the ALLWISE catalog obtained with the Wide-field Infrared Survey Explorer \citep{WISE}, the H$\alpha$ emission-line KISO survey \citep{KISO}, and the \textit{XMM-Newton} serendipitous sources catalog \citep{XMM}. We also investigated the dust distribution in this region based on the Planck 857\,GHz Survey image \citep{Planck1,Planck2}.

\begin{center}
\begin{table}[tph!]%
\caption{Catalogs and observations used in this study.}
\begin{center}
\begin{tabular}{ll}
\hline
\hline
Survey&Band/Channel\\
\hline
SDSS & $g$, $r$, $i$, $z$\\
2MASS&$J$, $H$, $K_s$\\
WISE&$W1$, $W2$, $W3$, $W4$\\
\textit{XMM}-Newton& 0.2-12\,keV \\
KISO& H$\alpha$\\
Planck&857\,GHz\\
\hline
\end{tabular}
\end{center}
\end{table}
\end{center}

\begin{figure*}[h!] 
        \centering
        \includegraphics[width=\textwidth]{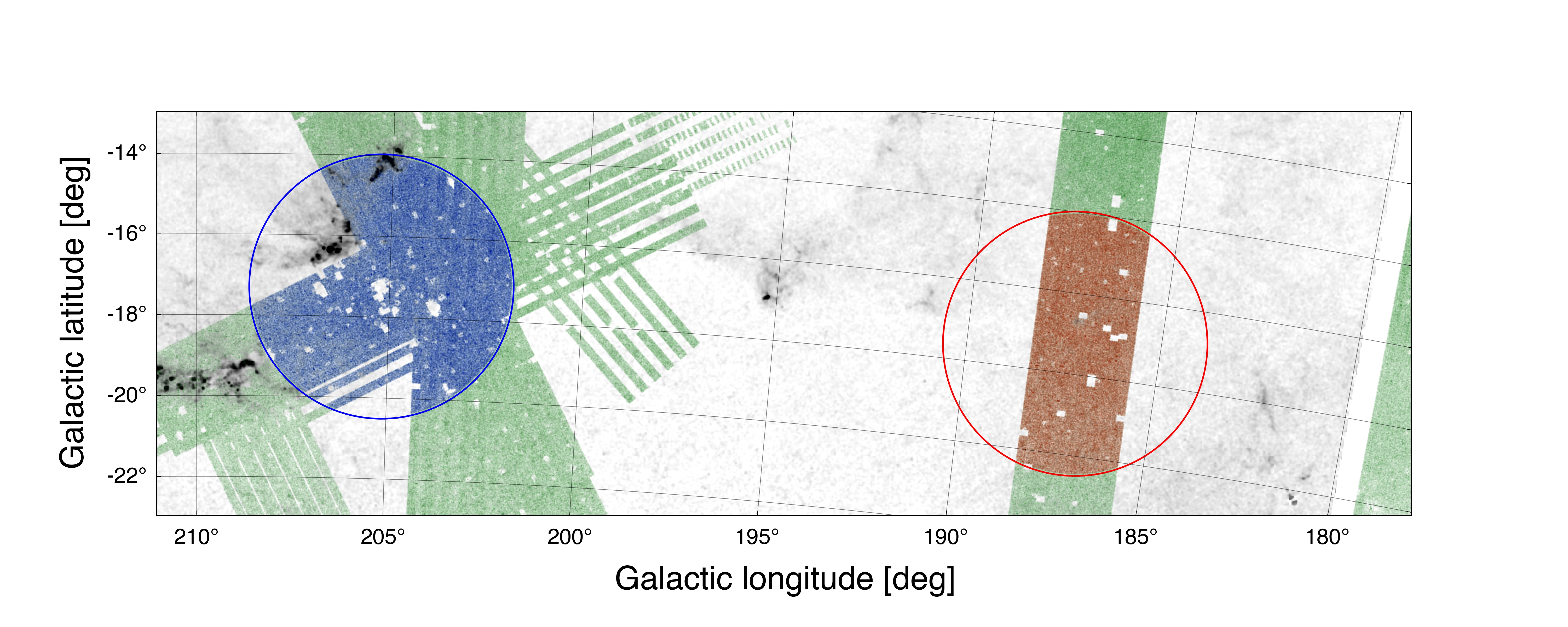}
        \caption{Science (blue) and control (red) fields plotted over the NICER dust extinction map by \cite{Lombardi2011}. Green represents the coverage of SDSS catalog. We note the areas without data, in particular around the Orion Belt bright stars.} \label{fig:fields}
\end{figure*}

\subsection{Control field}
Control fields provide an efficient and simple method to estimate excess stellar population statistics and luminosity functions. We selected the control field (CF) using the following criteria:
\begin{itemize}
\item Same size as the survey field,
\item Centered on the same Galactic latitude as the area of interest described above ($b$=$-17.24^{\circ}$) to minimize any enhancements in stellar surface density caused by the Milky Way structure,
\item Located in a region with low extinction, as reported in \cite{Lombardi2011} extinction map, to obtain the best characterization of the background population (Figure \ref{fig:fields}),
\item Covered by the SDSS\,DR12 catalog.
\end{itemize}
The selected CF is located  almost 30\degr\ away from the scientific field toward lower galactic longitudes and centered on $(l, b)=(187.0, -17.24)^{\circ}$. Owing to the peculiar specific coverage of the Sloan's survey away from the Galactic cap, this is the closest satisfactory field. Unfortunately, despite our best efforts, the control field is not extinction free and suffers from about 1 magnitude of visual extinction. Because we estimate the population size in the NIR, this is not critical because one magnitude of visual extinction is on the order of the extinction noise measurement in the NIR.  Figure \ref{fig:fields} shows the location of the science and control fields as blue and red circles, respectively, superimposed on NICER extinction map by \cite{Lombardi2011}. The green \textit{strips} correspond to the coverage of SDSS catalog, illustrating the incomplete coverage of optical data. The CF contains 167\,105 2MASS sources and 376\,846 SDSS sources, of which only 59\,546 sources have photometry available from both surveys.

\section{A new rich and large stellar group around $\epsilon$ Ori} \label{sec:results}

\subsection{Stellar surface density}\label{sec:surface_density}
\begin{figure}[h!] 
        \centering
      \includegraphics[width=\hsize]{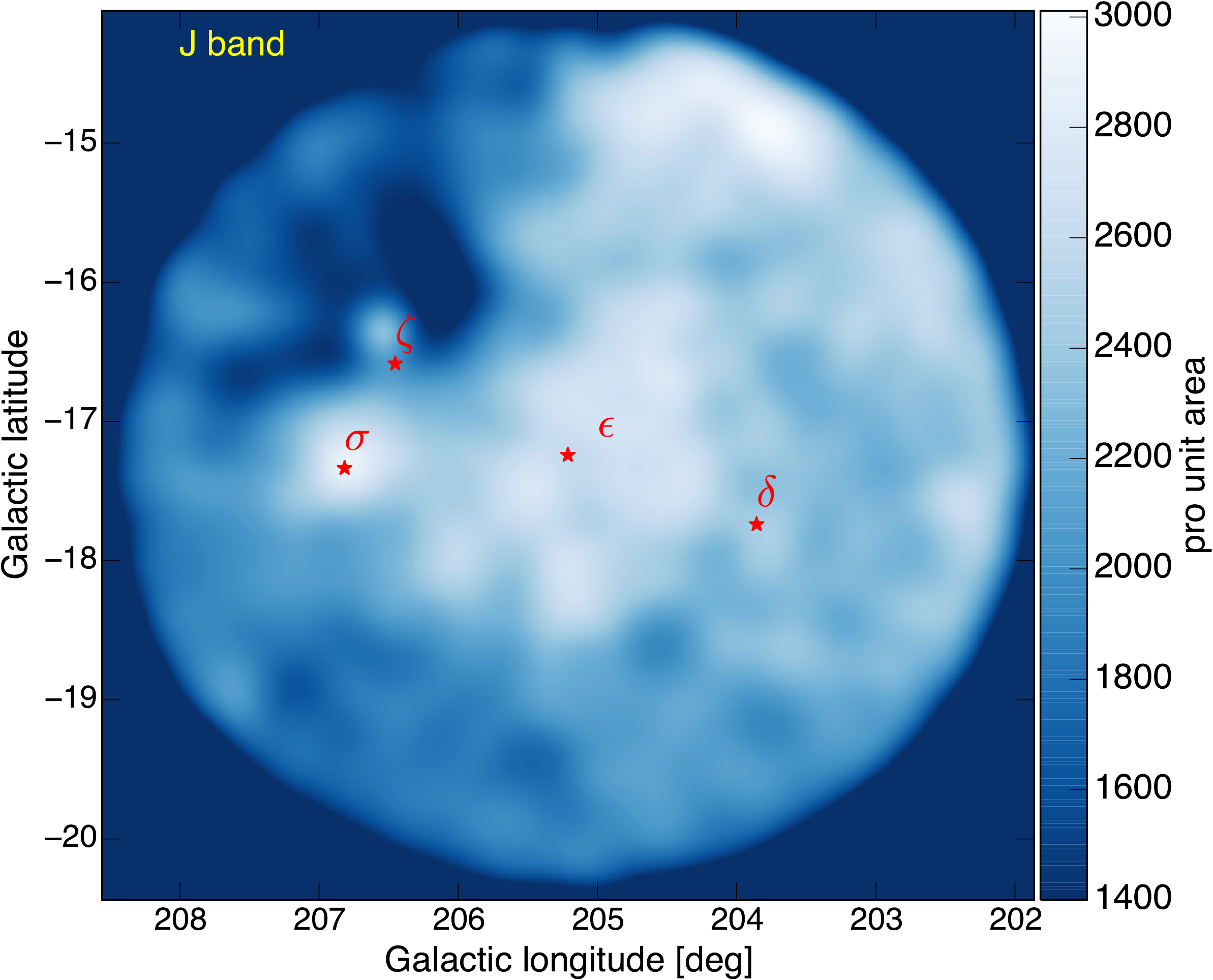}
                \caption{Surface densities calculated based on J-band 2MASS.  Asterisks denote the positions of the three Orion Belt stars and $\sigma$ Orionis. Similar results are obtained for the H and K bands.} \label{fig:J-KDE}
\end{figure}

The surface density of sources provides a simple first step to search for stellar groups. Figure~\ref{fig:J-KDE} shows the kernel density estimate (KDE) of the positions of sources in the 2MASS $J$ band within a magnitude cut of 15.3\,mag (0.2\,mag above the completeness limit to avoid the different level of photometric noise level for different 2MASS strips) and was obtained using a Gaussian kernel with a bandwidth of 10\arcmin. 

We investigated the KDE maps for all eight available bands with bandwidths ranging from 0.01\degr\ to 1\degr. The $ugriz$ density maps showed low-density areas around the bright OB stars of the Orion Belt and we interpret these as artifacts related to the presence of the bright stars, in particular the super-giants in Orion's Belt asterism. The near-IR 2MASS survey covers a broader dynamic range and is less affected by the presence of bright stars. Furthermore, the NIR is better suited to study the stellar spatial distribution thanks to its lower sensitivity to extinction and better sensitivity to low-mass stars that dominate the initial mass function (IMF). 

A clear overdensity is seen around $\sigma$\,Ori in the $JHK_s$ surface density maps, and $\zeta$\,Ori (Alnitak) in the $H$ and $K_s$ maps. These two density enhancements are associated with the well-known $\sigma$~Orionis and NGC~2024 stellar clusters \citep[e.g.,][]{Meyer+HandbookofSF}. Another density enhancement is clearly visible around $\epsilon$\,Ori (Alnilam) in the $J$ and $H$ maps. We measured the significance of overdensities by estimating a rms noise level from ten measurements around the main enhancement in Figure~\ref{fig:J-KDE}. The main overdensity toward the center of this figure, the main object of study throughout this work, is defined by a threshold of 7$\sigma$.

While the overdensity around $\sigma$ Orionis and NGC~2024 region are uniform and compact at the spatial scale probed by our analysis (10\arcmin), the density enhancement around $\epsilon$ Ori is much more extended and shows some internal structure. We also note the lack of any obvious density enhancement around $\delta$~Ori (Mintaka), in contrast with the result by \citealt{Caballero_Solano2008}, who investigated the population of young stars and brown dwarfs in a comparatively smaller region around Mintaka and Alnilam. It is still possible, however, that a potential enhancement around $\delta$~Ori is less significant than our 5$\sigma$ significance threshold. Finally, the KDE maps also show a surface density enhancement at almost all wavelengths in the northern part of the field, closer to the Galactic plane, which we tentatively interpret as the increasing stellar density produced by the Galactic structure and do not discuss further here. Finally, the Orion~B molecular clouds are clearly visible north of $\zeta$ and $\sigma$~Ori and appear as a region of lower density. Some of the properties of this enhancement are presented in Table \ref{tab:clusters}.

 \begin{figure*}[h!]
        \centering
                \includegraphics[width=\textwidth]{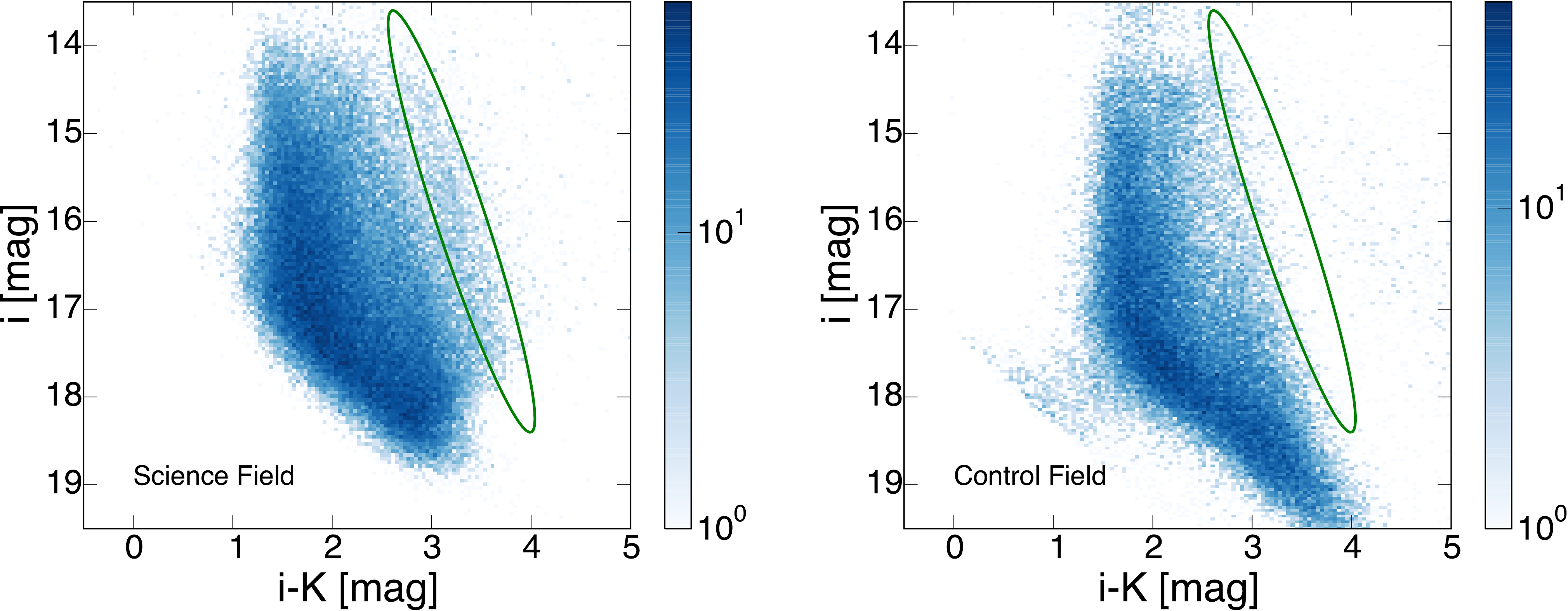}
                \caption{$(i, i-K_s)$ color-magnitude diagrams for all objects with 2MASS K$_s$ and SDSS $i$-band photometry. Left: sources in a 3\degr\ radius region centered on the B0\,Ib star, $\epsilon$ Orionis (the central star of Orion's Belt). Right: sources in the CF. The green ellipse plotted in each panel indicates the approximate position of the sequence in the science field.}
\label{fig:CMD-ivsi-K-ellipse.jpg}
\end{figure*}

\begin{center}
\begin{table}[h!]   
\caption{Properties of the main overdensity in Figure~\ref{fig:J-KDE}}
\begin{tabular}{p{3cm}p{3cm}p{2cm}}
\hline
\hline
Properties&&Value\\

\hline 
Significance& & $>$ 7$\sigma$\\
Size& Ellipse 1.7\degr $\times$2.2\degr & 2.94 \degr$^2$\\
Number of stars  & &2345\\
Total mass & & $\sim1200$ M$_\odot$\\
Surface density   && 800 stars/ \degr$^2$\\
   &for d$=$380 pc& 18 stars/pc$^2$\\
   &for d$=$250 pc& 41 stars/pc$^2$\\
Volume density &for d$=$380 pc& 8 stars/pc$^3$\\
&for d$=$250 pc& 17 stars/pc$^3$\\

\hline 
Position of peaks & $\sigma$-Ori &14$\sigma$\\ 
A &205.5\degr, -17.5\degr&11$\sigma$\\
B &204.5\degr, -16.8\degr&10$\sigma$\\
C &204.5\degr, -17.5\degr&9$\sigma$\\
D &205.2\degr, -18.2\degr&$8.5\sigma$\\

\hline
\normalsize
\end{tabular}
\label{tab:clusters}  
\end{table}
\end{center}

To estimate the size of the population included in the density enhancement around $\epsilon$~Ori, we estimate the number of sources falling into a 1.7\degr\,$\times$\,2.2\degr\ ellipse (7$\sigma$ threshold in overdensity in the J-band KDE map) encompassing it (9\,466 sources), and the number of sources included in an equal area located in the control field (7\,121 sources). The latter gives an estimate of the number of foreground and background sources that one can expect to find in that region of the galaxy. Subtraction of both values gives 2\,345 sources, which we use as an estimate for the size of the population producing the density enhancement. Repeating the count in 9 equal-area random positions within the CF leads to a dispersion of $\sim$215 sources.

\subsection{Color-magnitude and color-color diagrams of the survey}\label{sec:CMDs-CCDs}

Figure~\ref{fig:CMD-ivsi-K-ellipse.jpg} shows a $(i, i - K_s)$ color-magnitude diagram for the sample of sources in our survey (left panel) and CF (right panel). A rather dense sequence is clearly visible in the science field but not in the CF, suggesting the existence of a nearby young and rich population (indicated approximately by the green ellipse). The dispersion in color along the sequence, which is less than 1\,mag, is lower than the typical dispersion observed for very young clusters ($\leq$5~Myr, \citealt{Mayne2007}). 

The sequence is also clearly visible in other optical and NIR  color-magnitude diagrams for the science field, but is not present in the corresponding color-magnitude diagrams for the control field; this confirms that the population separates well photometrically from the field population across a broad range of wavelengths. A closer qualitative examination of the photometric properties of stars in the survey field can be carried out by comparing various color-color diagrams.

\subsection{Selection method}\label{sec:selection}

We took advantage of the clear separation of the sequence in various color-magnitude diagrams and the apparent absence of significant extinction to select the members of this new population. As in \citet{revisited2}, we applied the novel maximum-likelihood approach described in detail in \cite{Sarro_2014} to infer the membership of all the sources in our sample. This multidimensional probabilistic analysis offers the advantage of using multiple color-magnitude diagrams simultaneously and includes a statistically sound treatment of errors and censored data. The $u$ band was  excluded from the analysis because the observations are significantly shallower and extremely sensitive to interstellar extinction, excess emission related to accretion, and stellar activity. Table \ref{tab:MC} gives the list of all sources in our final catalog and includes the identification number, J2000 coordinates, $griz$ SDSS, and $JHK_S$ 2MASS magnitudes, as well as the membership probability computed as described above.

\begin{figure}[h]
\begin{center}
\includegraphics[width=\hsize]{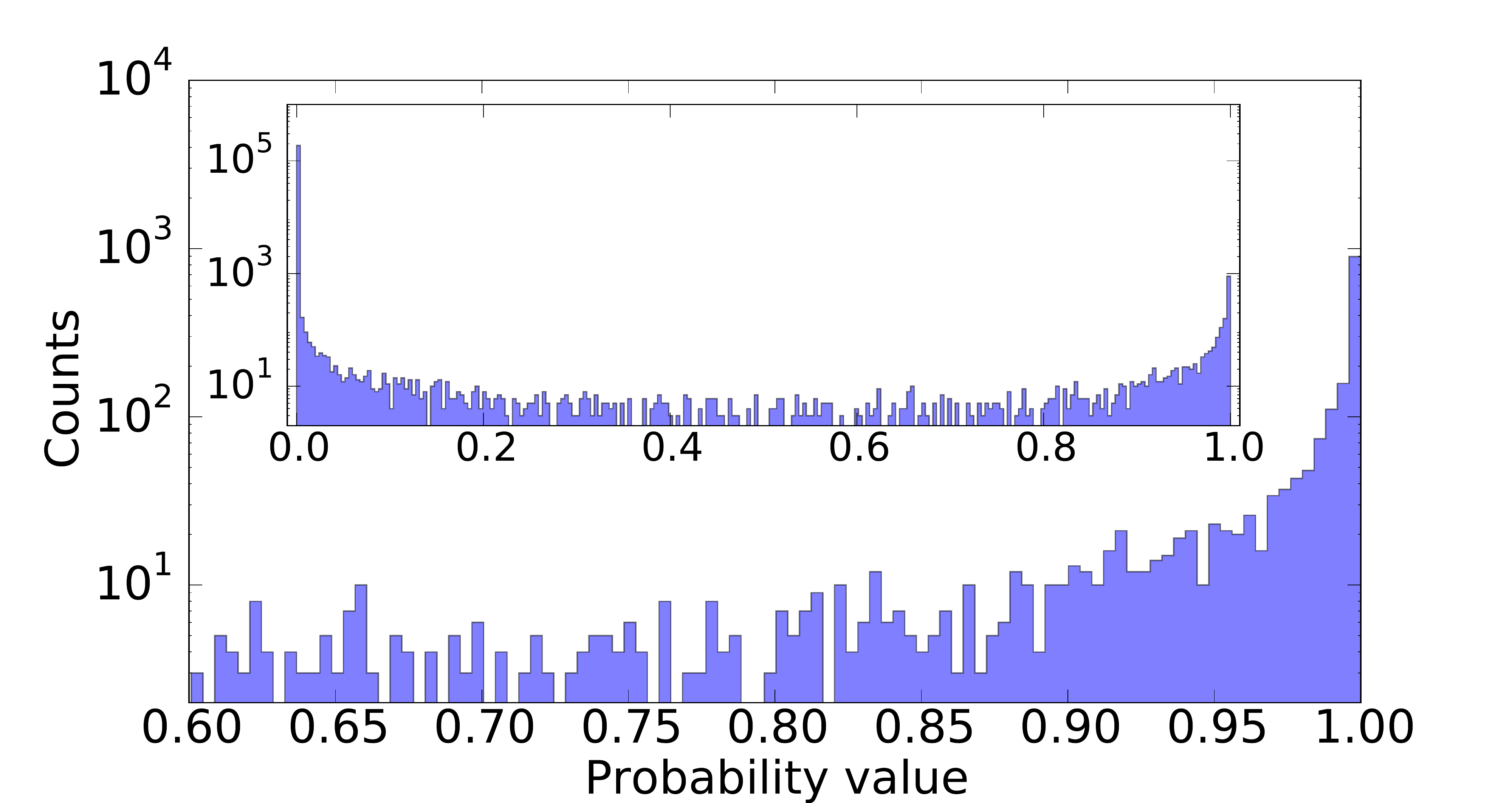}
\caption{Distribution of membership probabilities obtained for all sources (inner plot) and for sources with probability greater than 60\%. }
\label{fig:hist}
\end{center}
\end{figure}

The choice of a membership probability threshold is not trivial. Figure \ref{fig:hist} shows the distribution of membership probabilities for all the sources. It presents the typical bimodal distribution distribution comprised of a huge maximum around 0 and a smaller maximum around 1. Table \ref{tab:tresholds} shows the number of members for various threshold values.

\begin{center}
\begin{table}[tph]%
\caption{Number of members at different membership probability thresholds. }
\begin{center}
\begin{tabular}{lccccc}
\hline
\hline
prob. &0\%& 68\% &80\% &95\% &99.73\% \\
threshold&initial sample&&&\\
\hline
sample&189620&1956&1850&1494&783\\
size&&&&\\
\hline
\end{tabular}\label{tab:tresholds}
\end{center}
\end{table}
\end{center}

In an effort to be conservative, we select as members all the sources above a membership probability of 99.73\%, leading to a sample of 783 highly probable objects. 

\begin{figure*}[h!]       
       \centering
                \includegraphics[width=\textwidth]{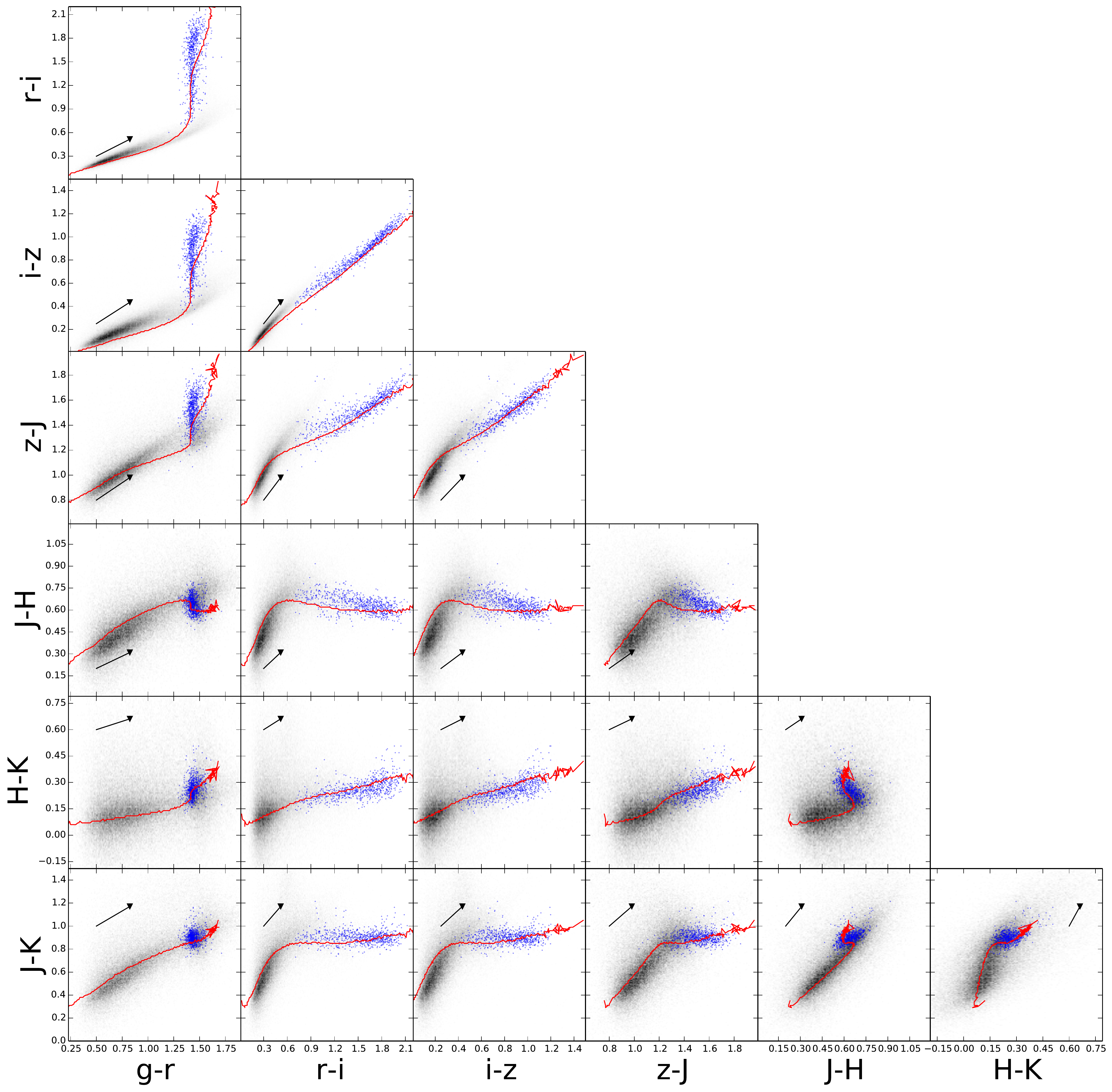}
               \caption{SDSS/2MASS color-color summarizing the photometric properties of the selected candidates (blue dots) compared to field sources (black dots). The color-color space of the selected sample coincides with that of unreddened M dwarfs. Also shown for each color-color space are the reddening vector for A$_\mathrm{V}$=1\,mag (black solid line) \citep{Cardelli1989} and the main sequence (red dashed line) from \citet{Covey2007}.}
                \label{fig:CCDs_Multi}
\end{figure*}

Figure~\ref{fig:CCDs_Multi} illustrates the efficiency of the multidimensional selection method used to compute membership probabilities. This figure shows a mosaic of color-color-diagrams constructed using the optical $griz$ and near-infrared $JHK$ colors of our input catalog. The high probability Orion Belt population (from hereafter, OBP) candidate members are overplotted with blue dots, and clearly separate from the field sources in most diagrams. The comparison of this positions of candidates with the intrinsic colors of luminosity V and III stars in the J - H versus H - Ks diagram from \cite{2MASS_colors}, shows, as expected, that the sample of OBP high probability members is made mostly of M stars.

This list of highly probable members is far from complete and suffers from several limitations. For example, there is no data close to the bright stars, contamination must still be present, and the different depths of the various optical and near-infrared data and the nonuniform spatial coverage of the data (in particular the SDSS data) biases the sample in some areas, and there are over specific luminosity ranges. This sample is nevertheless extremely useful to characterize the general properties of this new group and, in particular, its distance and age.

\section{Properties of the selected sample}\label{sec:properties}

\subsection{Spatial distribution}

\begin{figure}[h!]
        \centering
                \includegraphics[width=\hsize,origin=c]{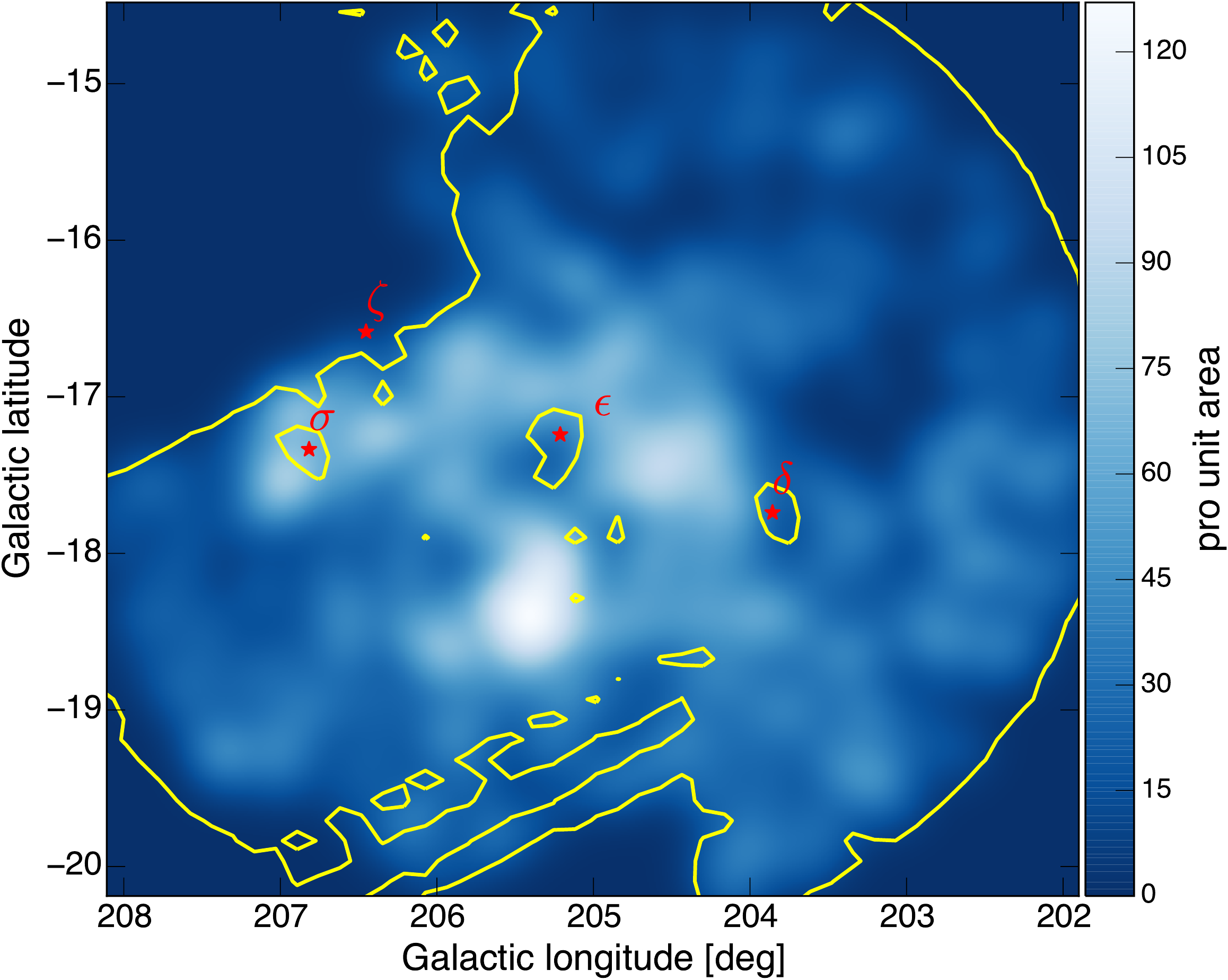}
                \caption{Surface density map of the 783 OBP candidate members. Yellow contours represent  source density from SDSS catalog. The map is affected by completeness near the  supergiants and $\sigma$\,Orionis ( indicated in red).}
\label{fig:8}
\end{figure}

Figure~\ref{fig:8} shows the 2D KDE of the spatial distribution of the 783 OBP candidate members computed using the same bandwidth as in Figure~\ref{fig:J-KDE}. The stellar density appears clearly lower in the close vicinity of the Belt supergiants ($\delta$, $\epsilon$, $\zeta$ Orionis) as a result of the incompleteness of the SDSS catalog near these bright stars, as illustrated by the yellow contours in Figure~\ref{fig:8}. A density enhancement is also visible around $\sigma$~Ori and suggests that our selection includes a few $\sigma$~Ori cluster members. As we see in Section~\ref{sec:comp_ngc1980_sigori}, the $\sigma$~Ori sequence is indeed very similar to that of the new population, but several pieces of evidence indicate that the two groups must be distinct. The corresponding contamination is nevertheless relatively low and concentrated around $\sigma$~Ori itself.
With these limitations in mind, we note that the most probable members seem to be located in an  0.5\degr\ wide ring that is roughly centered on $\epsilon$~Ori. The density in this ring is clearly not homogeneous, and a strong overdensity located south of $\epsilon$~Ori seems to dominate. The current data do not allow us to draw any further conclusions on the details of the spatial distribution of OBP members. More importantly, however, we retrieved a roughly similar structure to the J-band stellar density map (Figure~\ref{fig:J-KDE}) in a completely independent manner. This gives us confidence in the main result of this paper and the presence of a rich and fairly coeval stellar population toward the Orion Belt.     

To our knowledge, this coeval group does not correspond to any association previously identified in the literature. It is in particular more compact than what was defined as the Collinder~70 cluster (\citealt{Collinder}) and we propose that it is a distinct and new population of young stars.

\subsection{Age and distance estimates}

\subsubsection{Disk and accretor frequencies as an age diagnostic} 
We use the WISE and KISO $H\alpha$ surveys to study the disk and accretor frequencies among the sample of 783 high probability OBP members. The presence of a protoplanetary disk, which is probed by mid-infrared excesses in WISE, and intense accretion, which is probed by strong $H\alpha$ emission, provide clues on the age of a population. Protoplanetary disks that are responsible for the 3--12~$\mu$m excess emission revealed by WISE typically disperse over timescales of 5$\sim$10~Myr \citep{ribas2015}. Accretion of the circumstellar material onto the star stops being strong enough to produce sufficiently intense Balmer lines after a similar or shorter timescale. We find that only two stars within our sample have a counterpart in the KISO catalog (KISO A-0904 21 and KISO A-0903 163) and only 27 stars display mid-infrared excess in one or more WISE bands.\footnote{A detailed description of the mid-infrared excess analysis will be given in Kubiak et al., in preparation.} The KISO and WISE surveys sensitivities should encompass most of the luminosity range of our sample. These two small numbers suggest that the OBP members have cleared most of their protoplanetary disks and allow us to place a lower limit on the age around $\sim$5~Myr.

\subsubsection{Relation with the ISM and distance to the OBP}
\begin{figure}[h!]
        \centering
                \includegraphics[width=\hsize,origin=c]{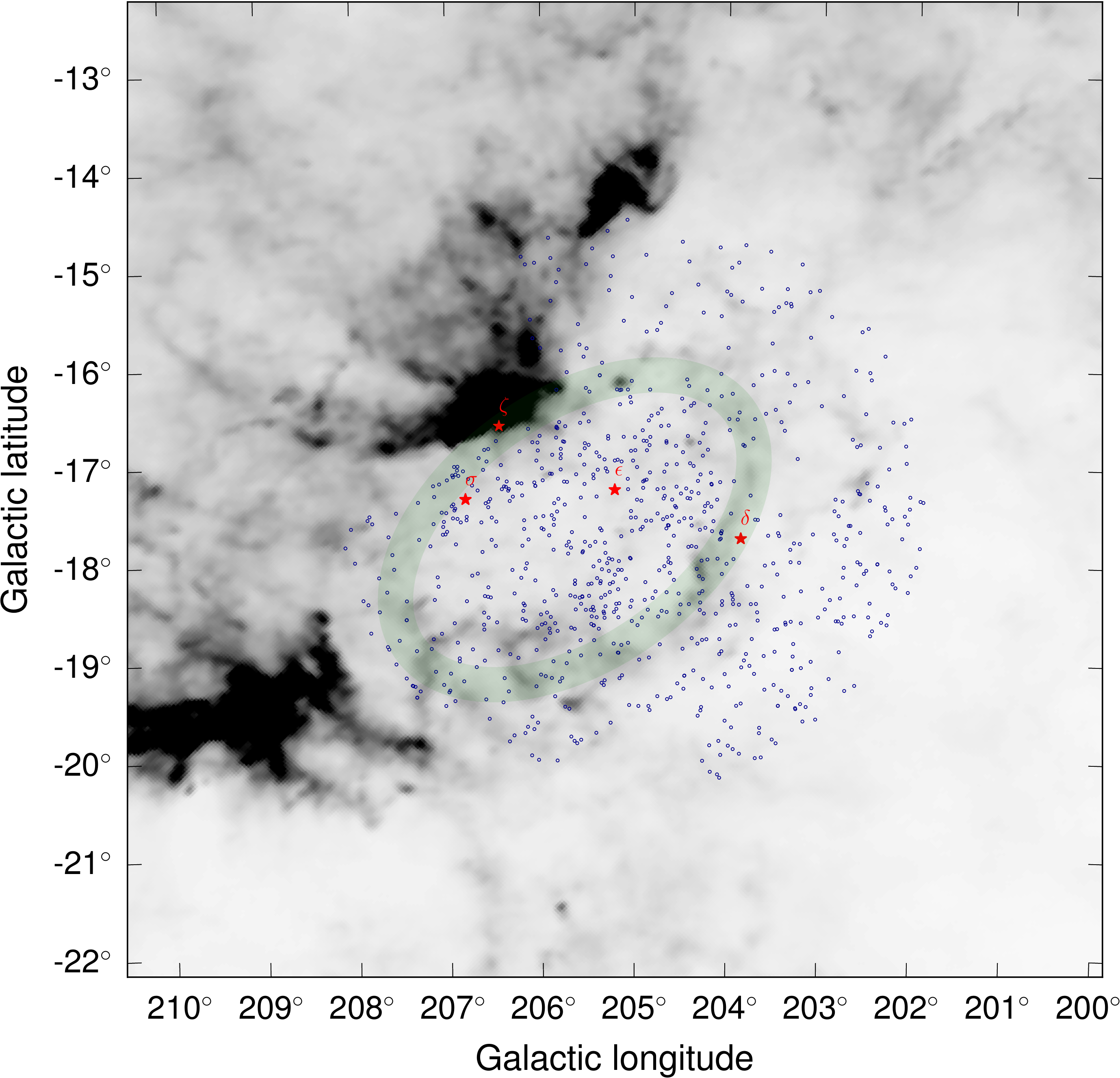}
                \caption{IC~434 region and surroundings as seen by Planck 857~GHz. The emission map shows a dust/gas shell \citep{Ochsendorf_2015} (indicated in light green) superimposed with the OPB (blue dots). We found no evidence for reddening in the OBP, implying that this stellar population must lie between Earth and the dust shell.}
                \label{fig:Planck}
\end{figure}

In the 857\,GHz Planck map, shown in Figure \ref{fig:Planck}, the Orion Belt population coincides, in part, with the IC~434 dust shell \citep{Ochsendorf_2015}. This gives us the opportunity to figure out which of the dust and the new population lies closer to Earth, as stars located behind dust clouds should appear reddened in color-color diagrams.

Figure \ref{fig:CCDs_Multi} includes a ($g-r$, $r-i$) color-color diagram for the 783 sources selected as the members of the Orion Belt population. These optical bands are particularly sensitive to extinction and a small amount of dust in the line of sight should be easily noticeable. The red line represents the intrinsic colors for main-sequence stars from \cite{Covey2007} and an extinction vector is indicated. As we can see, the members of the Orion Belt population create a narrow and compact sequence that is mostly unaffected by extinction and lies slightly left of the main sequence in the red. The amount of interstellar dust in the line of sight between Earth and the OBP must therefore be negligible, and the OBP stars must be in front of the IC~434 shell, which is itself located at a distance of $\sim$380~pc.

In an effort to obtain the minimum distance to the selected stars, we carry out a thought experiment. Knowing that our sample consists almost solely of M dwarfs, we estimate the distribution of brightness of M dwarfs for the different distances (from 100 to 400\,pc) and extinction values (A$\rm _v$ in range 0 \dots 1\,mag). Although this is not a simple exercise because we do not know the age of the population, and the luminosity of young M dwarfs drops rapidly after 10 Myr, we can make an educated guess that our sample cannot be closer than $\sim$250\,pc to the Sun.

\subsection{Comparison with other groups and clusters: $\sigma$-Ori and NGC1980 \label{sec:comp_ngc1980_sigori}}
\begin{figure*}[t!]
  \centering
    \includegraphics[width=.45\linewidth]{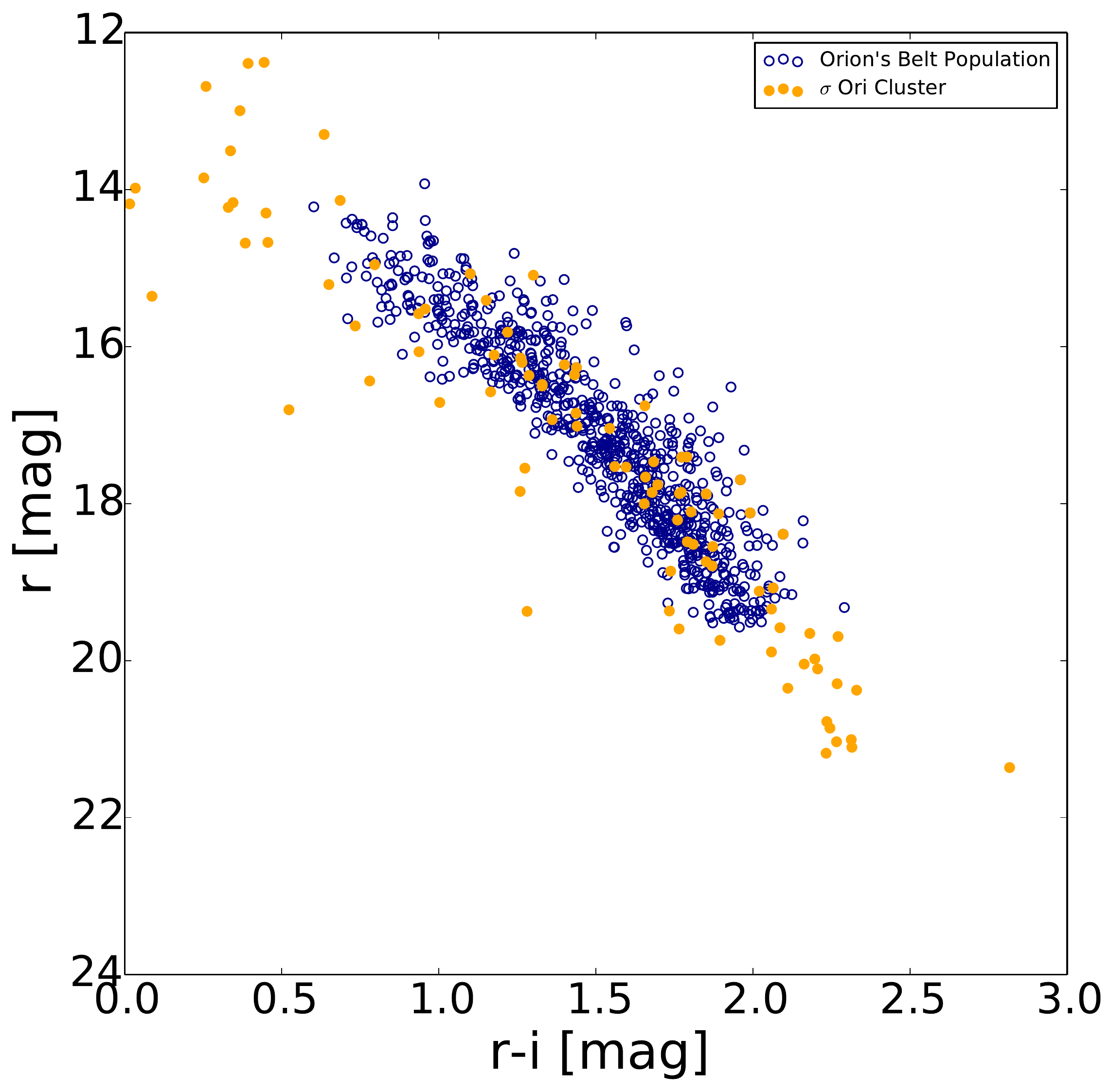}
    \includegraphics[width=.45\linewidth]{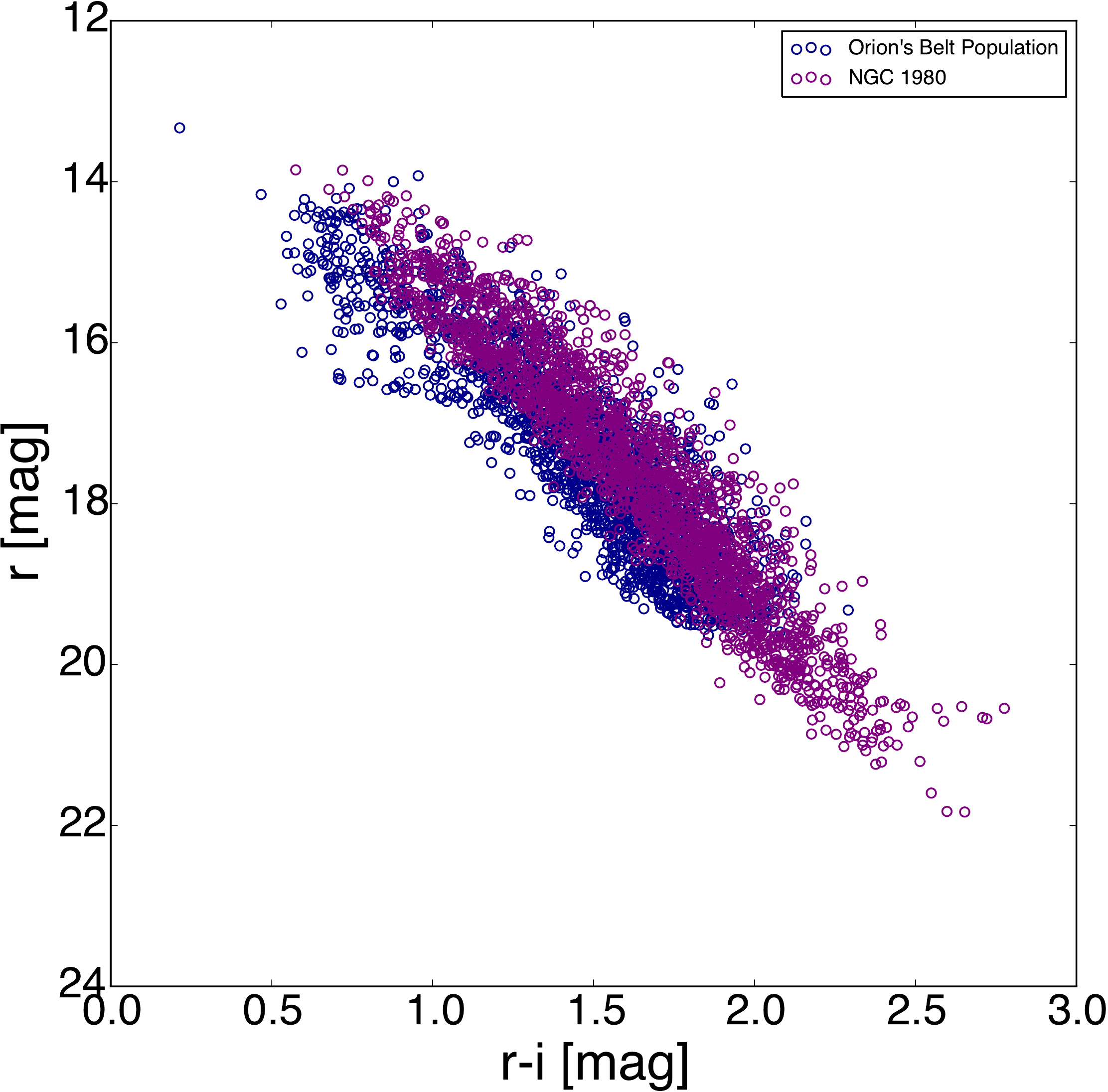}    
  \caption{The $r$ vs. $r-i$ color magnitude diagram for the OPB (blue circles) and $\sigma$~Orionis (orange circles) or NGC~1980 members (violet circles).} \label{fig:newfigure}
\end{figure*}

The comparison with empirical pre-main sequences of well-known young clusters in color-magnitude diagrams can provide clues to the age and distance of a young population. In Figure~\ref{fig:newfigure} we compare the sequence formed by the high probability OBP candidate members with those of  $\sigma$~Orionis \citep[3~Myr, 385~pc, and members list from][]{Caballero2007} and NGC1980 \citep[5--10~Myr, 400~pc, and members list from][]{revisited2}.

The figure shows that the OBP sequence is very similar to those of $\sigma$~Orionis and NGC1980. In fact, a cross-match between the three lists shows that 19 sources selected as OBP members were identified as $\sigma$~Orionis members by \citet{Caballero2007}, 52 sources selected as OBP members were identified as NGC1980 members in \citet{revisited2}, and this illustrates the complexity of isolating stellar groups in the Orion OBI region. 

The low disk and accretor frequencies among OBP members suggest that the OBP must be significantly older than the 3~Myr $\sigma$~Orionis, where as many as 35\% of T Tauri stars display mid-infrared excess related to the presence of a disk \citep{hernandez07}. For the two sequences to match, the OBP must therefore be closer to the Sun than $\sigma$~Orionis, in agreement with our conclusion based on the extinction toward the OBP members.

Figure~\ref{fig:newfigure} also shows that the OBP sequence is similar to that of NGC1980, but is slightly fainter by half a magnitude. If this difference is real, it could be the result of either a larger distance or an older age. In \citet{revisited2} of this series, we estimated an age of 5--10~Myr and distance of 380~pc for NGC1980. Since the comparison with $\sigma$~Orionis and the analysis of the extinction toward the OBP sources consistently imply a distance closer than 380~pc, we can rule out a significantly larger distance and conclude that the OBP must be older than NGC1980.

For the mass determination, we assumed an age of 5-10 Myr for the OPB since that was the best fit to the produced HR diagram. At this point, we should recall the assumptions and caveats in these estimates. The uncertainty in the determination of the population distance, together with the uncertainty in the age ($\sim$~5--10~Myr), are the largest contributors to the final error in the mass of each source; we assumed an uncertainty in the determination of the population distance of $\sim$380 pc, whilst
we believe it is an upper limit for distance determination in this case. Based on this distance, the least massive candidate members have masses of 0.05\,M$_{\odot}$. By assuming a normal IMF  (using both \citep{IMF_Chabrier} and \citep{IMF_Kroupa}) we estimate the total number of members as $\approx$ 2\,500.  This value should not be considered as the definitive size of the population, but an educated guess given the assumptions. Still, it is similar to the statistical estimate of source counts from the control field (2345$\pm$215), suggesting, although with strong caveats, that the population might have a normal IMF.

\section{Discussion}\label{sec:discussion}

We have found a population of about 2\,500 M stars (with 789 candidates with high probability) that are roughly coeval and extinction free and are distributed across $\sim3$ square degrees toward the Orion Belt asterism with an age of about 10 Myr. Photometry alone poorly constrains the distance to this population or its line-of-sight extent. This newly identified population can be as far as $\sim380$ pc (but in front of the Orion B cloud) or as close as 250 pc. In the closer case the OBP could be the low-mass counterpart to the well-known Orion supergiants at distances around 250 pc.

The new population, the OBP, is likely the low-mass counterpart of Blaauw's Ori OB Ib subgroup. Relevant to this discussion, \cite{Jeffries2006} performed radial velocity observations of low-mass stars toward a relatively large field toward $\sigma$ Ori and found two spatially superimposed components that are kinematically separated by 7 km/s in radial velocity and with different mean ages. These authors suggest an age of about 10 Myr for the older component (their ``group 1''), which has a mean radial velocity of 23.8 km/s. \cite{Jeffries2006} suggested that the older ``group 1'' was made by stars from the OB Ia subgroup, but the results in this paper suggest that this second component is most likely comprised of the stars in the Orion Belt population, or the OB Ib subgroup. Figure 2 of their paper further supports this statement as one can see how the field closer to the OBP (the NW field) is mostly dominated by stars belonging to the 23.8 km/s group. We cross-checked our list of targets against the sources in \cite{Jeffries2006} and although their study is centered on $\sigma$ Ori, we found that most of the matches with the OBP belong to the older group 1; albeit this finding also matches  group 2, which probably suggests that our selection method is not accurate enough to clearly separate the two different populations. Overall, our results reinforce the idea that overlapping populations at different evolutionary states and distances coexist along lines of sight toward the Orion clouds, as suggested in \cite{revisited1} and \cite{revisited2}.

\subsection{Is the Orion sequential star formation scenario in trouble?}

Blaauw's original idea of sequential star formation calls for a star formation event being directly responsible for the genesis of the next event. In Orion, it was proposed that the spatial-temporal sequence of events proceeded as follows \citep[e.g.,][]{Bally2008}:

\vspace{0.5cm}
\noindent
{\small Ia ({$\sim 12$ \rm{M}yr)} $\rightarrow$ Ib ($\sim 10$ \rm{M}yr) $\rightarrow$ Ic ($\sim 5-7$ \rm{M}yr) 
$\rightarrow$  Id ($\sim 1-3$ \rm{M}yr).} 
\vspace{0.1cm}

\noindent
In recent decades, evidence has been accumulating suggesting that this attractive scenario suffers from several shortcomings. \cite{Brown1994} found that subgroup Ib is younger than Ic and, to address the problem of an obvious break in the spatial-temporal sequence, these authors argued that the sequential star formation scenario is still plausible if the Ic population had moved from its putative birthplace closer to the Ia population; this move has yet to be quantified. Nevertheless, if the OBP is indeed the low-mass counterpart of Ib, then the results in this paper are in tension with \cite{Brown1994} as we find that the age of the OBP is similar to the canonical age of Ib (around 10 Myr), apparently solving the break in the spatial-temporal sequence.

Another problem for the sequential star formation scenario is the superposition of populations with different ages, as they do not easily fit a star formation sequence that covers about 100 pc from west to east. Evidence for such overlapping stellar populations has been accumulating in the literature \cite[e.g.,][]{Gomez1998,WarrenHesser1977,Jeffries2006,revisited1,revisited2}. For example, what event triggered the formation of the 1-3 Myr old $\sigma$ Ori cluster, seen along the same line of sight as the $\sim10$ Myr old OBP? Given that the Id subgroup is still forming stars and that Ia is too removed/old to be the trigger, one faces two options in a sequential star formation scenario: the trigger was either a) Ib or b) Ic, a subgroup closer to $\sigma$ Ori in age but not in projection. If a) then one needs to explain the roughly 7-8 Myr delay in the formation of $\sigma$ Ori. If b) one needs to explain the apparent break in spatial sequence (Ic is about 20 pc away from $\sigma$ Ori in projection, so probably in reality more). Option a) seems unlikely as Ib would have to trigger the formation of Ic to the southeast 5-7 Myr ago and the $\sigma$ Ori cluster 2-3 Myrs ago toward its background, as seen from Earth. Regarding option b), a possible solution to the break of the spatial-temporal sequence is to evoke that $\sigma$ Ori was formed elsewhere. This was recently suggested, in a different context by \citep{Ochsendorf_2015}. In the Ochsendorf-Tielens scenario $\sigma$ Ori was formed to the south (in galactic coordinates, see their Figure 1) of the Ic population and moved north toward the GS206-17+13 shell. It is hard to imagine how the feedback from Ic to the south would trigger the formation of the $\sigma$ Ori cluster and cause it to move north. In summary, neither option seems satisfactory.

\subsection{The Orion blue stream scenario}

Recently, \cite{streams} suggested a new scenario for the interpretation of the distribution of OB stars in the local neighborhood. In a reanalysis of the Hipparcos catalog, these authors found that the distribution of OB stars followed large-scale structures that are well-defined and elongated,  which they refer to as blue streams. The roughly constant width of the streams, together with a monotonic age sequence over hundreds of parsecs, suggests that they are the outcome of a large star formation event. They describe the existence of three streams in the local 500 pc neighborhood, one of these is the Orion stream, originating at the position of the Orion clouds and extending to regions as close to Earth as $\sim200$ pc, but likely even closer. This scenario imposes a well-defined age sequence as it assumes that young stars stream away from their place of birth, currently the Orion A and B molecular clouds. The further a population is from its birth place, the older it should be. Given the current position of the Sun in the Galaxy, the Orion stream appears projected along its length for an observer on Earth, which implies that stellar populations at different ages and distances should appear superposed. Because of the particular projection effect the Orion stream, this new scenario does not require a spatial sequence, unlike Blaauw's sequential scenario for Orion.

The new blue streams scenario appears to accommodate well the available observational data of the Orion star-forming region as a whole. As discussed above, there is plenty of evidence in the literature for superposition of populations with different ages along the direction to the Orion clouds. For example, the OBP fits well this new view of Orion as a stream projected along its length; a roughly 10 Myr old population is seen in projection toward a significantly younger $\sigma$ Ori, and an even younger NGC2024 cluster, still embedded in the Orion B cloud.  The OBP, we argue, is closer to Earth than $\sigma$ Ori and the cloud interacting with it via the H${\rm II}$ region. We also argue that there is some evidence that the OBP (Ib subgroup) is closer and older than NGC1980/NGC1981 (Ic subgroup), which is closer and older than the $\sigma$ Ori cluster. This age and distance relation is in good agreement with the blue streams scenario presented in \cite{streams}. Finally, the Ia subgroup and the 25 Ori cluster \citep[e.g.,][]{Briceno2008,Downes2014,Downes2015}, which are not addressed in this work, would also be part of the Orion stream. If older than the OBP, they should correspond to the nearest components of the Orion stream. But this remains to be confirmed, as the OBP could be older, hence nearer, and could be the low-mass counterpart to the well-known Orion supergiants at about 250 pc from Earth.

The streams scenario provides another advantage: it does not require that populations move substantially from each other, as proposed in \cite{Brown1994}, to solve the apparent break in the spatial-temporal sequence in Blaauw's scenario. In the streams scenario, the OB subgroups should have a space motion toward the same general direction, so a prediction of the streams scenario is that the proper motions between subgroups should be relatively small. In the streams scenario, Blaauw's OB subgroups could represent different components of the same stream with different ages and distances; these are all formed at about 400 pc by clouds long gone with the exception of subgroup Id, the ONC, embedded in Orion A cloud, and NGC2024 embedded in the Orion B cloud.

\subsection{Is the OBP the future of the ONC?}

Can the OBP be the evolved counterpart of an ONC-like cluster that was formed about 10 Myrs ago or is it an altogether different type of object? A striking property of the OBP population is that it is distributed over a large area of the sky and its low stellar density is very different from other well-known stellar clusters in Orion, such as the ONC, $\sigma$, $\lambda$, or $\iota$ Ori clusters. For example, the average stellar surface density in the Orion nebula cluster ($\sim200$ stars/pc$^{2}$; \cite{Hillenbrand_Hartmann_1998}) is about an order of magnitude higher than that of the OBP.  The volume density of stars in the core of the ONC  ($2-3\times10^4$ stars/pc$^{3}$, \cite{Hillenbrand_Hartmann_1998}) is about three orders of magnitude higher than that of the OBP. 

Remarkably, both the ONC and the OBP have a similar number of stars. Assuming that the OBP is not a gravitationally bound population, and that it is expanding freely since it got rid of its parental molecular cloud early in its formation, it would have taken about 5 Myr for the OBP\ to expand from 2 pc to 7 pc radius at about 1 km/s expansion velocity, or 10 Myr for 0.5 km/s velocity. These rough estimates are not implausible according to models of an ONC-type cluster expanding after gas removal \citep[e.g.,][]{Kroupa_2001}, and so the possibility that the OBP might represent an evolved ONC cannot be discarded with current data. High-resolution spectroscopic observations or accurate proper-motions measurements are needed for a more quantitative answer to this question; such measurements and observations do not exist at the moment.

\section{Summary}\label{sec:summary}

In order to find the spatial extension of the foreground stellar population to the Orion A cloud found in \cite{revisited1} and \cite{revisited2}, and to investigate the relation between Blaauw's OB\,Ic and Ib subgroups, we analyzed a circular area with a radius of 3$^\circ$ centered on $\epsilon$ Orionis (HD 37128, B0Ib), covering the Orion Belt region. The main results of this investigation are as follows:

\begin{itemize}
\item We found two large stellar overdensities in the Orion Belt region: one centered on the well-known $\sigma$-Ori cluster, and a new, richer but more extended overdensity close to $\epsilon$ Ori. We compared the stellar density in the surveyed region with a control field and estimated an upper limit for the size of the new overdensity of about 2\,345$\pm$215 sources.

\item Optical and near-IR color-magnitude diagrams reveal a well-defined sequence above the Galactic field, which is suggestive of a large young stellar population that is approximately coeval and not affected by interstellar extinction.  We used a new statistical multiband technique to select objects associated with the sequence detected in the color-magnitude diagram, and compiled a catalog of 783 probable members. Essentially, all of these objects have the colors of M stars. The selected sources are close, in projection, to $\epsilon$ Ori, but distributed in a roughly elliptical region ($1^\circ \times 3^\circ $) showing spatial substructure.

\item This new population, that we call the Orion Belt population, is likely the low-mass counter part to the Ori OB Ib subgroup. We found a negligible amount of bona fide young stellar objects in the Orion Belt population (less than 2 \% for all available youth tracers (\textit{XMM-Newton}, KISO, and WISE surveys). This allows us to infer the minimum age of the cluster to be $\sim5$\,Myr. We estimate an age of about $\sim10$\,Myr for the OBP.

\item We do not find evidence for an interaction between the selected members and the clouds, which together with the overall absence of extinction suggests that the new population lies in the foreground of Orion~B. We estimate the distance to this newly identified population to be between  $\sim$250 and  $\sim$380\,pc.

\item Although our results do not rule out Blaauw's sequential star formation scenario for Orion, we argue that the current available evidence is shifting against it. We find, instead, that the blue stream scenario proposed in \cite{streams} provides a better framework on which one can explain the Orion star formation region as a whole. 

\item We speculate that the Orion Belt population could represent the evolved counterpart of a Orion nebula-like cluster. At least high-resolution spectroscopic data would be needed to make a more solid statement about the origin of this newly identified population. 

\end{itemize}

Finally, although we argue that the OBP fits the blue stream scenario best, we caution that independent work is needed to confirm the existence of the blue streams. Nevertheless, giving the tantalizing proximity and youth of the new stellar population presented in this work, there is a need for a dedicated spectral and dynamic characterization of the OBP. This population could become a benchmark region for future searches of brown dwarfs and planetary mass objects and the low-mass end of the IMF, as well circumstellar disk evolution and planet formation. The final ESA Gaia catalog, to be released around 2023, will include much if not all of the OBP candidates presented in this work, and will be able to shed much light on the origin of the OBP, the existence and role of the Orion blue dtream, and star formation in Orion.

\begin{appendix}
\section{final catalog}\label{App:MasCat} 

Table \ref{tab:MC} provides the photometric data for the candidate members of the Orion Belt population; it contains the name of each star, right ascension and declination, $girz$ and $JHK_s$ band magnitudes from the SDSS and 2MASS catalogs with their associated uncertainties. This table is available in its entirety in a machine-readable form in the online journal. A portion is shown here for guidance regarding its form and content.

\begin{sidewaystable}
\caption{The final catalog\label{tab:MC}}           
\centering        
\begin{tabular}{ccccccccccc}    
\hline\hline       
ID&R.A.&Dec&g&i&r&z&J&H&K$\rm _S$&Prob.\\
&(J2000)&(J2000)&(mag)&(mag)&(mag)&(mag)&(mag)&(mag)&(mag)&(\%)\\
\hline
1&83.928031&-4.0978&20.132$\pm$0.019&16.709$\pm$0.0060&18.624$\pm$0.01&15.673$\pm$0.0070&13.979$\pm$0.035&13.359$\pm$0.043&13.072$\pm$0.03&99.9\\
2&83.949248&1.887384&18.378$\pm$0.0070&15.415$\pm$0.0050&16.873$\pm$0.0060&14.628$\pm$0.0050&13.205$\pm$0.027&12.595$\pm$0.026&12.323$\pm$0.023&99.9\\
3&82.444109&1.523527&16.621$\pm$0.0040&14.383$\pm$0.0040&15.225$\pm$0.0030&13.859$\pm$0.0040&12.481$\pm$0.024&11.827$\pm$0.024&11.611$\pm$0.021&99.8\\
4&83.125498&-3.991311&17.657$\pm$0.0060&14.701$\pm$0.0050&16.195$\pm$0.0050&13.857$\pm$0.0050&12.405$\pm$0.033&11.835$\pm$0.03&11.548$\pm$0.024&99.9\\
5&83.126417&-3.990177&20.353$\pm$0.023&16.903$\pm$0.0060&18.806$\pm$0.011&15.816$\pm$0.0070&14.207$\pm$0.044&13.577$\pm$0.046&13.29$\pm$0.051&99.9\\
...&...&...&...&...&...&...&...&...&...&...\\
779&83.738131&-0.802611&16.332$\pm$0.0040&14.062$\pm$0.0040&14.919$\pm$0.0040&13.534$\pm$0.0040&12.207$\pm$0.022&11.478$\pm$0.03&11.3$\pm$0.023&99.95\\
780&84.089487&-1.031828&19.386$\pm$0.013&16.321$\pm$0.0050&17.952$\pm$0.0070&15.358$\pm$0.0050&13.823$\pm$0.024&13.253$\pm$0.037&12.996$\pm$0.03&100.0\\
781&83.70118&-1.146224&18.743$\pm$0.0090&15.836$\pm$0.0040&17.316$\pm$0.0050&15.049$\pm$0.0050&13.57$\pm$0.039&12.979$\pm$0.044&12.684$\pm$0.039&99.8\\
782&83.826168&-1.014437&18.465$\pm$0.0080&15.595$\pm$0.0040&16.992$\pm$0.0050&14.811$\pm$0.0050&13.35$\pm$0.026&12.623$\pm$0.032&12.41$\pm$0.026&100.0\\
783&83.856769&-1.048855&18.014$\pm$0.0070&14.921$\pm$0.0040&16.602$\pm$0.0050&14.01$\pm$0.0040&12.45$\pm$0.024&11.839$\pm$0.033&11.526$\pm$0.019&99.9\\
\hline
\end{tabular}

\end{sidewaystable}

\end{appendix}

\begin{acknowledgements}
J. Alves acknowledges travel support from the ESAC Faculty council. H. Bouy is funded by the Ram\'on y Cajal fellowship program number RYC-2009-04497.  This research has made use of the VizieR catalog access tool, CDS, Strasbourg, France. The original description of the VizieR service was published in \cite{Vizier}. This research has made use of "Aladin Sky atlas" developed at CDS, Strasbourg Observatory, France \cite{Aladin1} and \cite{Aladin2}. This research has made use of Topcat \citep[\url{http://www.starlink.ac.uk/topcat/},][]{2005ASPC..347...29T}. This research made use of Astropy, a community-developed core Python package for Astronomy \citep{2013A&A...558A..33A}.

\end{acknowledgements}


\bibliographystyle{aa}
\bibliography{Orion.bib}

\end{document}